\documentclass[aps,prb,twocolumn,showpacs]{revtex4}
\usepackage{graphicx}

\begin{document}

\title{Pulling gold nanowires with a hydrogen clamp}

\author{Sz.~Csonka, A.~Halbritter, and G.~Mih\'aly}
\affiliation{Electron Transport Research Group of the Hungarian
Academy of Sciences and Department of Physics, Budapest University
of Technology and Economics, 1111 Budapest, Hungary}

\date{\today}

\begin{abstract}
In this paper an experimental study of the interaction of hydrogen
molecules with gold nanowires is presented. Our results show, that
chains of Au atoms can also be pulled in hydrogen environment,
however in this case the conductance of the chain is strongly
reduced compared to the perfect transmission of pure Au chains.
The comparison of our experiments with recent theoretical
prediction for the hydrogen welding of Au nanowires implies that a
hydrogen molecule can even be incorporated in the gold
nanocontact, and this hydrogen clamp is strong enough to pull a
chain of gold atoms.
\end{abstract}

\pacs{73.63.Rt, 73.23.-b, 85.65.+h}

\maketitle

\section{Introduction}

The behavior of gold nanojunctions have been widely investigated
in the past years. Due to its inertness, gold is an easy-to-use
material, and Au nanojunctions can even be studied under ambient
conditions with a low risk for contamination. This monovalent
metal was found to be a perfect candidate for studying the quantum
nature of conductance in a single-atom contact. It was shown by
various experimental and theoretical methods that a monoatomic Au
junction has a single conductance channel with perfect
transmission.\cite{Agrait2003} Surprising was the discovery that
Au has a special property of forming monoatomic chains during the
elongation of a single-atom contact.\cite{Yanson1998}

Recently, another remarkable property of gold was proposed:
various theoretical simulations have pointed out that - contrary
to the well-known inertness of macroscopic Au surfaces - {\it gold
nanostructures} are strongly reactive constructions of
nature.\cite{Bahn2002,Barnett2004} This proposal is supported by
experimental studies on the catalytic properties of Au
nanoclusters.\cite{Zhai2004} The possibility for the chemical
interaction of gold nanowires with adsorbants was initially
advised in order to explain the anomalously large inter-gold
distances in electron microscope images of Au
nanochains.\cite{Bahn2002,Novaes2003,Legoas2004,Skorodumova2003}
In this paper we present an experimental study on the chemical
interaction of gold nanochains with the simplest molecule, H$_2$.
We compare our results to recent theoretical calculations
investigating the possible configurations, where hydrogen is
chemically bound to the Au nanowire.\cite{Barnett2004}

\begin{figure}[t!]
\centering
\includegraphics[width=\columnwidth]{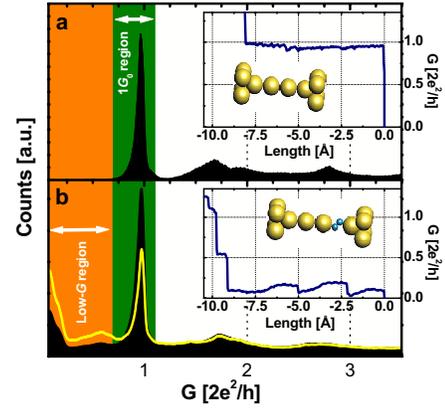}
\caption{\it Panel (a) is the conductance histogram of pure gold
in high vacuum, exhibiting a sharp peak in the $1G_0$ region. This
peak arises due to flat conductance plateaus at $1G_0$
corresponding to single atom Au contacts or monoatomic Au wires
(the latter is shown in the inset, an interatomic Au-Au distance
in the chain is $2.5\,$\AA). The area graph in Panel (b) is the
histogram of gold in hydrogen environment, showing the appearance
of new atomic configurations in the low-$G$ region. The inset
shows an example for a trace with a well-structured periodic
behavior in the low-conductance region, which is attributed to the
pulling of a gold chain with an incorporated H$_2$ molecule. The
line graph in panel (b) is the histogram for the hydrogen affected
traces (see text).} \label{CorrHist.fig}
\end{figure}

Our measurements were performed on high-purity gold samples with
the mechanically controllable break junction (MCBJ) technique
under cryogenic circumstances. We have investigated the
characteristic behavior of the nanocontacts by acquiring a large
number of conductance traces during the repeated separation of the
electrodes, and collecting the data into conductance
histograms.\cite{Circumstancesnote} A typical conductance
histogram measured in ultra high vacuum is presented in
Fig.~\ref{CorrHist.fig}a. The most pronounced feature is a sharp
peak at the quantum conductance unit, G$_0=2e^2/h$. This peak
arises due to flat conductance plateaus at $1G_0$ corresponding to
single atom Au contacts or monoatomic Au wires (see inset) both
having a single conductance channel with almost perfect
transmission.\cite{Agrait2003} The lower edge of the histogram
shows that nanocontacts with lower conductance can also occur down
to $0.7$\,G$_0$, but below that value the histogram has no counts.
In hydrogen environment, however, a low-conductance tail arises in
the histogram [area graph in Fig.~\ref{CorrHist.fig}b], that is
the conductance of the nanojunctions can take any value below the
quantum unit. The growth of the tail in the histogram is
attributed to the appearance of new, low-conductance atomic
arrangements in the hydrogen environment. Usually a peak near
$0.5$\,G$_0$ is also superpositioned on the low-conductance tail
with varying amplitude, which was in the focus of our previous
work.\cite{Csonka2003} The object of this study is to understand
the microscopic nature of the new hydrogen-related atomic
configurations through the statistical analysis of the conductance
traces. The emphasis is put on the whole low-conductance tail, and
the results are insensitive to the weight of the peak at
$0.5$\,G$_0$. First we show with the investigation of plateau's
length histograms that gold nanocontacts in hydrogen environment
are generally exhibiting the formation of atomic gold chains,
however, the interaction with hydrogen can strongly reduce the
transmission of the chain. Next, we demonstrate a unique feature,
which is observed on a portion of the conductance traces showing a
well-structured periodic behavior. An example for such traces is
already presented in the inset of Fig.~\ref{CorrHist.fig}b. Based
on recent theoretical simulations we attribute this phenomenon to
the pulling of gold nanowires with an incorporated hydrogen
molecule.

Our measurements are in close relation to recent studies on the
interaction of hydrogen molecules with platinum and palladium
nanocontacts.\cite{Smit2002,Csonka2004} In both cases the
histogram in high vacuum shows a pronounced peak at $\sim
G=1.5\,G_0$, which is attributed to the conductance of single atom
Pt and Pd contacts. The interaction with hydrogen, however,
completely suppresses this peak, and new, hydrogen-related peaks
grow. In the case of platinum the vibrational modes detected by
nonlinear $I-V$ characteristics gave clear evidence that a
molecular hydrogen bridge is formed between the platinum
electrodes, which has perfect transmission through a single
conductance channel. In palladium the transmission is reduced due
to the dissolution of H atoms in the Pd electrodes.

In gold the novel phenomena due to the interaction with hydrogen
is reflected by the growth of the low-conductance tail in the
histogram, but the presence of hydrogen does not destroy the sharp
peak at the quantum unit, which is attributed to single atom Au
contacts or monoatomic Au chains. It implies that in gold the
interaction with hydrogen can coexist with the characteristic
behavior of pure gold junctions.

\section{Chain formation in hydrogen environment}

As it was presented in Fig.~\ref{CorrHist.fig}, hydrogen induces
new low-conductance configurations. Further on it will be
demonstrated by the statistical analysis of the conductance traces
that the observed new configurations are also related to atomic
gold chains.

To make a quantitative analysis possible, we define two
conductance regions, as shown by the shadowed areas in
Fig.~\ref{CorrHist.fig}. The first region (``$1G_0$ region'') is
identified with the interval where the peak at $G=1\,G_0$ grows in
the conductance histogram of pure gold, $G/G_0 \in [0.7, 1.1]$.
The second, so-called ``low-$G$'' region is the interval, where
the new hydrogen-related configurations arise in the histogram,
$G/G_0 \in [0.05, 0.7]$.

We found that even in hydrogen surrounding a lot of traces show
the behavior of pure gold, exhibiting plateaus near $1$\,G$_0$
without any data points in the low-$G$ region. The histogram for
these curves completely resembles the one taken in the absence of
hydrogen. Next we investigate the other portion of the traces that
do show low-$G$ configurations. We shall call a trace
``hydrogen-affected'' if it stays for a reasonable length of
$>0.25\,$\AA\ in the low-$G$ region.

With these definitions we find that typically $\sim50\%$ of the
traces is hydrogen-affected. The histogram for this portion of the
traces is plotted by the line graph in Fig.~\ref{CorrHist.fig}b.
This histogram can be directly compared to the one for the whole
data set (area graph), as both of them are normalized to the
number of traces included. Naturally, for the hydrogen-affected
curves the weight in the low-$G$ region is significantly larger.
At $G>1.1\,G_0$ the two histograms precisely fall onto each other,
whereas the peak near the conductance quantum is significantly
suppressed for the hydrogen-affected traces. It implies, that the
appearance of the new low-$G$ configurations does not affect the
evolution of the conductance of larger junctions, but it strongly
influences the behavior of single atom contacts or atomic chains.
A deeper insight into this correlation can be obtained by
investigating the distribution of the lengths of the traces in the
above defined two regions.

\subsection{Plateaus' length analysis}
\label{PLHanal}

\begin{figure}[t!]
\centering
\includegraphics[width=\columnwidth]{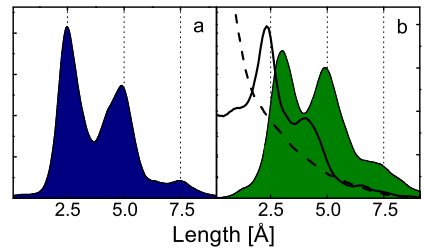}
\caption{\it Plateaus' length histograms for gold junctions in
hydrogen environment. Panel (a) shows the plateaus' length
histogram for the traces that are not affected by hydrogen. Panel
(b) contains plateaus' length histograms for the hydrogen-affected
traces  ($\sim 50\%$ of all curves). The line graph shows the
distribution of the lengths in the $1G_0$ region, the dashed line
shows the distribution of the lengths in the low-$G$ region,
whereas the area graph presents the distribution of the summed
lengths of the low-$G$ and $1G_0$ regions.} \label{PlatLength.fig}
\end{figure}

Figure~\ref{PlatLength.fig}a presents the distribution of the
lengths in the $1G_0$ region for those traces that are not
affected by the hydrogen. This so-called plateaus' length
histogram shows an equidistant peak structure, which is the proof
for the formation of chains of gold atoms, as it was shown by
Yanson et al.\cite{Yanson1998} The peaks in this histogram
correspond to the rupture of chains with different number of gold
atoms included, and the distance between the peaks corresponds to
the interatomic Au-Au distance in the chain. This plateaus' length
histogram is identical to the one taken before admitting hydrogen
to the junction, thus it can be regarded as a reference histogram
showing the formation of pure gold atomic chains. The peaks in
this histogram are also used to calibrate the length scale, as in
a pure gold atomic chain the interatomic distance was measured to
be $2.5\,$\AA.\cite{Untiedt2002}

\begin{figure}[t!]
\centering
\includegraphics[width=0.9\columnwidth]{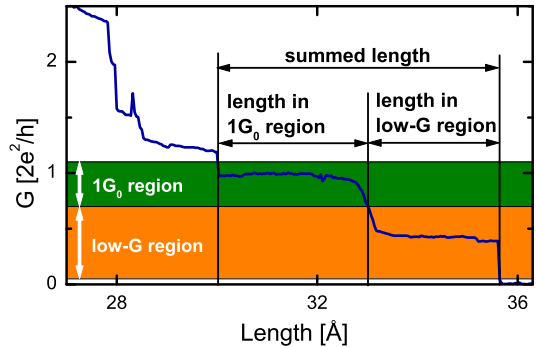}
\caption{\it The figure shows the determination of the lengths in
the different conductance regions for an individual conductance
trace. The length of a trace is measured by the number of data
points in a certain region. The summed length means the length in
the entire $1G_0$ and low-$G$ region, i.e. $G/G_0 \in [0.05,
1.1]$.} \label{LengthMeas.fig}
\end{figure}

Next we investigate how this chain formation is influenced by the
interaction with hydrogen. In the case of pure gold the
conductance of the last atomic configuration before the breakage
always shows a flat plateau in the close vicinity of $1G_0$ (see
the inset in Fig.~\ref{CorrHist.fig}a), and accordingly the
plateaus' length histogram is built from the distribution of the
lengths in the $1G_0$ region. In the case of the
``hydrogen-affected'' curves, however, the conductance can take
any value below the quantum unit, thus it makes sense to study the
distribution of the lengths both in the $1G_0$ and in the low-$G$
region. The determination of these lengths is demonstrated in
Fig.~\ref{LengthMeas.fig}.

The proper analysis of the distribution of the plateau's length
gives distinctive information about the possible microscopic
choreographies of the rupture. For instance a strong interaction
with hydrogen might completely destroy the chain formation, which
would result in the disappearance of the equidistant peak
structure in the plateau's length histogram, and only the first
peak would survive. As another possibility, one can consider that
perfectly transmitting chains are formed just like in pure
environment, and the new low-conductance configurations are only
appearing after the rupture of these chains. In this case the
distribution of the plateau's length in the $1G_0$ region should
show the same equidistant peak structure as the reference
histogram in Fig.~\ref{PlatLength.fig}a. In contrast, the
measurement shows that for the hydrogen affected traces the
plateau's length histogram for the $1G_0$ region is strongly
distorted compared to the reference histogram (solid line in
Fig.~\ref{PlatLength.fig}b). A plateau's length histogram can also
be plotted for the low-$G$ configurations, exhibiting a completely
featureless distribution (dashed line in
Fig.~\ref{PlatLength.fig}b). Surprisingly, the distribution of the
{\it summed length} of the low-$G$ and $1G_0$ region [area graph
in Fig.~\ref{PlatLength.fig}b] shows the well-defined peak
structure just like the reference histogram
[Fig.~\ref{PlatLength.fig}a]. (For the definition of the summed
length see Fig.~\ref{LengthMeas.fig}.) Observing peaks at the same
equidistant positions as in the case of pure gold demonstrates
that the chain formation is preserved in hydrogen environment. But
contrary to pure gold, the peaks appear in the plateaus' length
histogram of the entire 1G$_0$ and low-G region which also has an
important consequence that the new low-conductance configurations
are not arising {\it after} the rupture of the chain, but both the
$1G_0$ and the low-$G$ configurations are a part of the chain
formation. This can be explained as follows: an atomic gold chain
is formed just like in pure environment, but during this process
at a certain point the hydrogen binds to the chain, which results
in the strong reduction of the transmission. An individual trace
corresponding to this process is shown in
Fig.~\ref{LengthMeas.fig}. It is noted that the interaction with
H$_2$ strongly reduces the conductance of the chain, but the
atomic periodicity agrees with that of pure Au chains.

\begin{figure}[t!]
\centering
\includegraphics[width=\columnwidth]{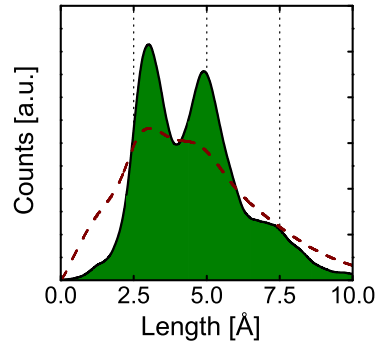}
\caption{\it Analysis of the plateaus' length histogram for the
summed length region. The area graph  presents the measured
distribution of the summed lengths of the low-$G$ and $1G_0$
regions. The dashed line shows the calculated distribution of the
summed length which means the convolution of the plateaus' length
histograms for the low-G region and  1G$_0$ region assuming that
the lengths in these two regions are independent. The appearance
of the peaks demonstrates the correlations indicating the chain
formation in the entire low-$G$ and $1G_0$ regions.}
\label{Convolution.fig}
\end{figure}

As the length of the entire 1G$_0$ and low-G regions  follows a
well-defined, periodic distribution, the lengths in its two
subregions must be correlated. This correlation is demonstrated in
Fig.~\ref{Convolution.fig}. If the length of the configurations in
the low-G and 1G$_0$ region were independent, the distribution of
the summed length could be expressed simply as the convolution of
the distribution functions of the two subregions, which is shown
by the dashed line. The calculated convolution curve assuming the
independence of the subregions only gives the rough tendency of
the measured one (area graph), but it can not explain the presence
of the peaks. It nicely demonstrates again that the low-G region
and 1G$_0$ region must be considered together and the entire
$G/G_0 \in [0.05, 1.1]$ is the physically relevant conductance
region for the chain formation process in hydrogen environment.

\begin{figure}[t!]
\centering
\includegraphics[width=0.75\columnwidth]{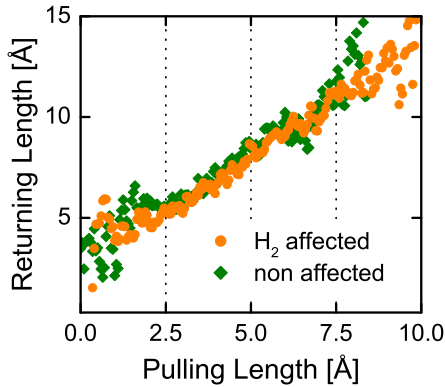}
\caption{\it The average returning length as a function of the
pulling length for the hydrogen-affected traces (circles) and the
non-affected traces (diamonds). The pulling length is the length
of the plateau in the $[0.05,1.1]$ region before break, whereas
the returning length is the length measured from point of breakage
back to point, where a contact with $G>0.05$\,G$_0$ is
reestablished.} \label{ReturnLength.fig}
\end{figure}

Further confirmation for the chain formation process for the
hydrogen affected curves can be obtained by the statistical
analysis of the returning lengths following the idea of Yanson et
al.\cite{Yanson1998} If an atomic chain is being broken the atoms
in the chain collapse back to the electrodes, thus in order to
reestablish the contact the electrodes must be pushed towards each
other by almost the same distance as the length of the chain was.
This means that the relation between the length of the plateau
before the break and the so-called returning length is
approximately 1:1. Fig.~\ref{ReturnLength.fig} shows that this
relation holds not only for the traces that are non-affected by
the hydrogen but also for the hydrogen affected curves.

\subsection{New class of chains}
\label{NewClass}

\begin{figure}
\centering
\includegraphics[width=\columnwidth]{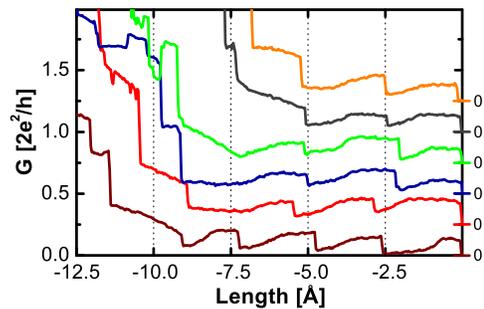}
\caption{\it Conductance traces showing a well-defined periodic
structure in the low-$G$ region. The bottom curve is on true
scale, while the other curves are artificially shifted on the
conductance axis for the sake of visibility (the corresponding
zeros are signed on the right axis). The interatomic distance in
the pure gold chain is indicated by grid lines. Approximately
$5\%$ of the traces shows this special behavior.}
\label{LongCurves.fig}
\end{figure}

\begin{figure}
\centering
\includegraphics[width=\columnwidth]{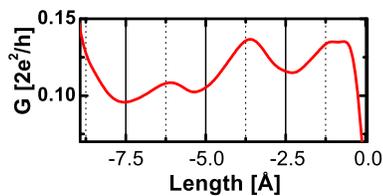}
\caption{\it The average of all the traces that have a length
$>8.8 \AA$ in the low-$G$ region. The averaging was done as a
function of the length measured backward from the point of
breakage.} \label{AvrCurve.fig}
\end{figure}

In previous studies solely the analysis of the distribution of the
plateaus' length was applied to demonstrate the chain formation
process and to determine the atomic periodicity of the chains, as
generally the chain formation is not visible on the shape of
individual conductance traces. In hydrogen environment, however,
we have found that a part of the traces ($\sim5\%$ of all curves)
exhibit a well structured periodic behavior in the low-$G$ region,
as demonstrated by the curves in Fig.~\ref{LongCurves.fig}. These
traces show periods with positive slope, where the conductance is
significantly increasing as the electrodes are pulled apart. At
the end of these periods the conductance frequently jumps to the
initial value, and the same behavior is repeated several times.
Generally the conductance of these periodic sections is situated
in the bottom of the low-G region, that is in the interval $G/G_0
\in [0.05, 0.3]$ (which will be called as  ``tail region''). The
peak-to-peak amplitude of these oscillations can be as large as
$0.2G_0$, which frequently exceeds the mean value of the
conductance. The grid lines of the figure are drawn with a
separation corresponding to the interatomic distance of the pure
gold chains, $2.5$\AA. It is apparent that the length of these
periods is close to the interatomic gold-gold distance in a
monoatomic chain, which implies that these special conductance
traces are also related to the chain formation.

 To check the statistical
relevance of the behavior demonstrated by the curves in
Fig.~\ref{LongCurves.fig}, we have calculated an average
conductance curve [Fig.~\ref{AvrCurve.fig}]. In order to determine
the average behavior along several interatomic distances, we have
restricted the averaging for those traces that stay for a length
of $>8.8\,$\AA\ in the low-$G$ region. The average conductance
curve also shows the oscillatory behavior with a period coinciding
with the interatomic gold-gold distance, showing that the well
structured periodic behavior of the traces in the low-$G$ region
is a substantial feature of gold nanojunctions breaking in
hydrogen environment. We attribute this behavior to a unique type
of contact breakage, corresponding to a special microscopic
process, which will be discussed in Sec.~\ref{comparison}.

\subsection{Comparison of the two types of chains}

The plateaus' length analysis discussed in Sec.~\ref{PLHanal} has
shown that gold nanocontacts in hydrogen environment are generally
showing the formation of atomic gold chains, however the
interaction with hydrogen can strongly reduce the transmission of
the chain. In sections~\ref{NewClass} Figs.~\ref{LongCurves.fig}
and \ref{AvrCurve.fig} demonstrated a well-structured periodic
behavior, which is observed only on a portion of the traces, and
which is also related to some kind of chain formation process. In
the latter case the chain formation is not just statistically
detectable, but it is seen on the shape of the individual
conductance curves. In the following part it will be analyzed how
these two phenomena are related to each other.

\begin{figure}[t!]
\centering
\includegraphics[width=.95\columnwidth]{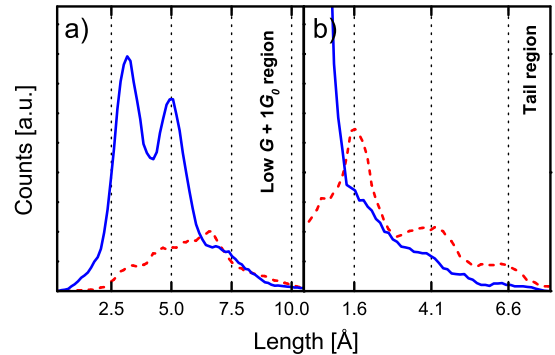}
\caption{\it Plateaus' length analysis for different types of
hydrogen affected traces. The plateaus' length histograms shown by
the dashed/solid lines are built from the traces that have
longer/shorter than 30\% section with positive plateaus' slope. In
panel (a) the distribution of the summed length in the entire
$1G_0$ and low-$G$ region is plotted, while panel (b) shows the
distribution of the lengths in the tail region, $G/G_0 \in [0.05,
0.3]$} \label{PlatLength2.fig}
\end{figure}

It is to be emphasized that not only the periodicity but also the
pronounced positive slope is an interesting peculiar property of
the conductance curves presented in Fig.~\ref{LongCurves.fig}. The
positive slope of a conductance trace means that the transmission
of a certain atomic configuration is increasing despite of the
fact that the electrodes are pulled apart and so the average
distance between the atoms is increasing. Such an unusual behavior
is not observed in pure gold junctions. Surveying the individual
conductance traces in hydrogen surrounding one by one we found
that practically only the curves presenting the well-defined
periodic structure exhibit positive slope. Therefore, this is a
distinctive feature, which can be used to statistically separate
the special periodic traces from the rest of the hydrogen affected
curves. We found that the periodic traces can be selected with
good certainty by sorting out those curves, for which the length
of the parts with positive slope in the low-G region is larger
than $30\%$ of the the whole length in the low-G region. With this
analysis one can plot the plateaus' length histograms separately
for the well structured periodic traces, and for the rest of the
curves, as shown in Fig.~\ref{PlatLength2.fig}. The dashed curves
are the plateaus length histograms for the well-structured
periodic traces with pronounced positive slope, whereas the solid
lines show plateaus length histograms for the rest of the hydrogen
affected traces.

Panel (a) in Fig.~\ref{PlatLength2.fig}  presents the plateaus'
length histograms for the summed length in the low-G and 1G$_0$
region. The traces without pronounced positive slope (solid line)
show the same behavior, which was already demonstrated in
Fig.~\ref{PlatLength.fig}: an atomic chain formation process
taking place in the entire conductance region of $G/G_0=[0.05,
1.1]$. Contrary, the traces with well-structured periodic behavior
do not show this feature, the plateaus length histogram for these
curves is completely featureless.

As the special periodic behavior basically appears in the tail
region, it is interesting to plot the plateaus' length histograms
for the tail region as well (Fig.~\ref{PlatLength2.fig}b). In this
case the distribution of the lengths for the periodic traces shows
equidistant peaks separated by a distance of 2.5 \AA\, whereas the
distribution of the lengths for the rest of the curves is
completely featureless.

This analysis shows that two different types of hydrogen related
chain formation processes can be distinguished. The first one is
seen for the majority of the hydrogen affected traces, during the
pulling of a chain at a random point the conductance switches from
the 1\,G$_0$ value to the low-$G$ region, but the entire length
remains the same as in pure gold. This is demonstrated by the
solid line in Fig.~\ref{PlatLength2.fig}a. The second one is only
observed for a portion of the traces ($\sim 5\%$), the conductance
of the chain drops to the tail region where a well-structured
periodic behavior is visible on every subsequent gold-gold
interatomic distance (see the dashed curve in
Fig.~\ref{PlatLength2.fig}). In the next part the possible
microscopic background for these two processes is discussed.

\section{Towards the microscopic picture: comparison with theory}
\label{comparison}

Further steps towards the microscopic understanding of the
observed phenomena can be done by comparing our experimental
results with recent computer simulations on the interaction of
hydrogen molecules with gold nanocontacts. Barnett et al.\ have
found that the basically inert gold surface becomes highly
reactive with hydrogen at nanoscale due to the relativistic
corrections in the electronic structure.\cite{Barnett2004} Various
atomic configurations were studied, including a hydrogen molecule
weakly bound to the side of a gold chain, a H$_2$ molecule
incorporated in the gold chain, and dissociated H atoms sitting in
the chain. This study has shown that a H$_2$ molecule incorporated
in the gold chain is a robust configuration with high binding
energy, which can be reached without any barrier. All these
configurations have conductances in the range of $0-0.25\,G_0$,
which coincides with the interval, where the strongest weight is
observed in the low-G region of the hydrogen-affected conductance
histogram [Fig.~\ref{CorrHist.fig}b].

\begin{figure}[t!]
\centering
\includegraphics[width=\columnwidth]{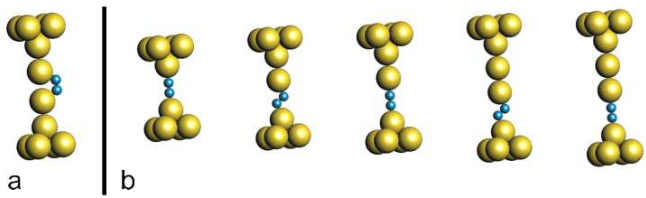}
\caption{\it Illustration for the weakly bound hydrogen molecule
(a) and the mechanism of pulling an atomic gold chain with a
hydrogen molecule (b).} \label{H2clamp.fig}
\end{figure}

The configuration where the hydrogen molecule is weakly bound to
the side of the chain (see Fig.~\ref{H2clamp.fig}a) might give an
important contribution to the general behavior demonstrated by the
plateau's length histogram [Sec.~\ref{PLHanal},
Fig.~\ref{PlatLength.fig}]. The formation of atomic gold chains is
preserved, but the chain gets distorted due to the binding of the
H$_2$ molecule and thus its transmission is reduced.

On the other hand the configuration, where a hydrogen molecule is
incorporated in the chain provides a reasonable explanation for
the pronounced positive slope of a part of the conductance traces
in our measurements [Sec.~\ref{NewClass},
Fig.~\ref{LongCurves.fig}]. The simulations have shown that the
molecule can be oriented in different angles with respect to the
contact axis. The perpendicular configuration has a smaller
conductance of $\simeq 0.1\,G_0$, but as the electrodes are pulled
apart and the molecule turns towards the parallel configuration
the conductance grows up to $\simeq 0.25\,G_0$. During this
process the force acting between the molecule and the gold
nanojunction reaches values as high as $\simeq 0.5$nN, and before
the complete rupture it can even grow to $\simeq
1.4$nN.\cite{Barnett2004} Note that these forces are in the same
range as the ones required for pulling monoatomic gold chains
($0.7-1.5\,$nN).\cite{Rubio2001}

Based on these considerations we give an explicit proposal for the
well-defined periodic behavior of the conductance traces
demonstrated in Fig.~\ref{LongCurves.fig}. We suggest that as a
gold nanojunction is pulled apart, a hydrogen molecule gets
incorporated in the chain. During further retraction of the
electrodes this hydrogen clamp does not break, but it is strong
enough to pull further gold atoms into the chain. Once a new gold
atom jumps into the chain the molecule gets in a tilted position
with smaller conductance, but as the contact is pulled the
molecule straightens out and thus the conductance increases until
the point where the next atom jumps into the chain. An
illustration of this process is shown in Fig.~\ref{H2clamp.fig}b.

 We have also performed $I-V$ curve measurements in order to
determine the vibrational energies of the molecule in the chain
similarly to Ref.~\onlinecite{Smit2002}. We have found, that in
the low-$G$ range the nonlinearity due to quantum interference
phenomena \cite{Ludoph1999,Untiedt2000} is so high that the
vibrational spectrum could not be resolved.

\section{Conclusions}

In conclusion, we have shown that hydrogen molecules can strongly
interact with gold nanojunctions, which is reflected by the
appearance of new, low-conductance atomic configurations. By the
analysis of plateaus' length histograms we have demonstrated that
these low-$G$ configurations are not established {\it after} the
formation of monoatomic gold chains, but they are a natural part
of the chain formation process. In addition we have shown, that
the conductance traces frequently show a well-defined periodic
behavior with positive slope in the low-$G$ region. This
observation is suggestive for a specific chain formation process:
a hydrogen molecule gets incorporated in the gold nanojunction,
and this hydrogen clamp is strong enough to pull a monoatomic gold
chain from the electrodes.

This work has been supported by the Hungarian research funds OTKA
F049330, TS049881. A.~Halbritter is a grantee of the Bolyai
J\'anos Scholarship.


\begin{thebibliography}{16}
\bibitem[{Eve()}]{Circumstancesnote}
\bibinfo{note}{The measurements were performed at typical temperatures of
$10-20\,$K and bias voltages of $100-120\,$mV, either in cryogenic
vacuum, or with a small amount ($1\,\mu$mole -- $1\,$mmole) of
H$_2$ admitted into the sample space. The histograms were built
from $20000$ conductance traces.}
\end{thebibliography}
\end{document}